\documentclass{kluwer} 
\newdisplay{guess}{Conjecture}
\usepackage{epsfig}
\begin{document}
\begin{article}
\begin{opening} 
\title{The star formation history as a function of type:\\ 
constraints
from galaxy counts\thanks{rocca@iap.fr, 
fioc@zardoz.gsfc.nasa.gov}} 
\author{Brigitte \surname{Rocca-Volmerange (1)} and Michel \surname{Fioc (2)}} 
\runningauthor{Brigitte Rocca and Michel Fioc}
\runningtitle{Star formation history from counts}
\institute{(1) Institut d'Astrophysique de Paris 98bis Bd Arago F-75014 Paris\\
(2) NASA/Goddard Space Flight Center, Greenbelt, MD 20771, USA}
\date{September 22, 1999}
\begin{abstract}
Deep galaxy counts are among the best constraints on the cosmic star
formation history (SFH) of galaxies. 
Using various tracers, the evolution of the star formation activity may
now be followed on a wide range of redshifts ($0 \leq z \leq 4$)
covering most of the history of the Universe. Two incompatible interpretations
of the observations are currently competing.
After applying
star formation rate (SFR) conversion 
factors to the CFRS, H$\alpha$ or ISO samples, many authors
conclude to a strong increase 
($\simeq$ a factor 10) of the SFR from $z=0$ to $z=1$. 
They also find some evidence for a peak at $z\simeq 1$
and for a rapid decrease at higher redshifts. 
On the other side, the Hawaii deep surveys 
favor only a mild increase between $z=0$ and $1$ (Cowie et al., 1996, 1999).

In this paper, we tackle this problem from
the point of view of the modelist of the spectral evolution of
galaxies. To understand the reason for these discrepant
interpretations, we consider three classes of galaxies: E/S0
(``early-type''); Sa--Sbc (``intermediate-type''); 
Sc--Sd, irregulars and bursting dwarfs (``late-type'').
We use the new version of our evolutionary synthesis
code, \textsc{P\'egase} (Fioc and Rocca-Volmerange, 2000, in preparation), which takes
into account metallicity and dust effects. 
The main results are: i) Late-type galaxies contribute significantly
to the local SFR, especially bursting dwarfs (Fioc and Rocca-Volmerange, 1999). Because of that,
the \emph{cosmic} SFR can not decrease by a factor $10$ from $z=0$ to $1$.
This is in agreement with Cowie et al., 1999 's result.
ii) The SFR of intermediate-type galaxies has strongly decreased since
$z=1$. Though the decrease is less than what find Lilly et al., 1996, 
this suggests that the CFRS and H$\alpha$~surveys are dominated 
by such bright early spirals. The limits in surface brightness and magnitudes
of the observed samples may be the main reason for this selection.
iii) The contribution of 
early-type galaxies increases rapidly from $z=1$ to their redshift of 
formation ($\geq 2$-3  for 
cosmological reasons). Their intense star formation rates at high-$z$ 
give strong constraints on early ionization phases, primeval populations 
or metal enrichments .
\end{abstract}
\end{opening} 

\section{Introduction} 
Long before the supernovae were used to explore the distant Universe,
faint galaxy counts were
known to be very sensitive to the cosmological parameters and to the
redshift of formation. Robust conclusions were derived on the
inconsistency with the data of a deceleration parameter $q_0=0.5$ and
a null cosmological constant (Yoshii and Peterson, 1991; Koo,
1990; Guiderdoni and Rocca-Volmerange, 1990). A flat universe
$\Omega_0=1$ could be saved only by either invoking a non-zero cosmological
constant $\Lambda_0$ (Fukugita et al., 1990) or by number evolution
(Rocca-Volmerange and Guiderdoni, 1990). No plausible change in star
formation parameters might alter these conclusions. So that we adopt 
hereafter the best values of the cosmological parameters for
our analysis of the cosmic star formation rate (SFR) in the Universe,
leaving to further studies, the sensitivity of results to this choice.

A large variety of redshift and photometric surveys have been used to
trace the SFH. Between $0\leq z \leq 1$, a ten-fold increase of the SFR
(Madau et al., 1996) has been derived from the Canada-France Redshift
Survey (CFRS, Lilly et al., 1996) from rest-frame 2800\,\AA\ computed
by interpolation or, at low-$z$, extrapolation of observed optical and
near-infrared data. Similar results were claimed using 
the H$\alpha$ surveys (Gallego et
al., 1995, Tresse et al., 1998) and the
ISO/CFRS data (Flores et al., 1998). On the other hand,
the complete redshift and photometric (from U' to K bands) 
surveys of galaxies
observed by Cowie et al., 1996, 1999 conclude to a milder evolution
$SFR\propto (1+z)^{1.5}$ on the same range of redshift and to a higher
local SFR, in agreement with the results of the FOCA2000 $z$-survey
(Treyer et al. 1998; Sullivan et al. 1999). 

Multiwavelength galaxy counts might help to solve this controversy. 
Interpretations of faint galaxy counts were recently proposed by 
Pozzetti et al., (1998) and, using the spectral evolution model 
\textsc{P\'egase}, by Fioc \& Rocca-Volmerange (1999a). The latter
notably identified a significant population of bursting dwarf galaxies 
in the FOCA 2000\AA\ photometric survey (Armand \& Milliard,
1994), whose contribution in the optical and near-infrared is
however much smaller than that due to the bulk of normal galaxies. 
In the optical-NIR domain, 
 galaxy counts are attributed to various populations of 
the Hubble sequence, distributed according to the observed 
local luminosity functions by spectral types 
(Marzke et al, 1994, 
Heyl et al, 1997). 
Each evolution scenario, mainly constrained
by local and low $z$~ observations, corresponds to a  
star formation law. The 9 types computed are hereafter gathered in 
 three groups:
Ellipticals-S0, spirals Sa-Sbc, Sc-Irr-dwarfs. In
the following we shortly recall the evolutionary modeling of galaxy
counts with P\'EGASE. Then we present the global SFH with a flat
increase for 0$ \leq z \leq $1, lightly shallower that Cowie's et al.,
1999. At
higher redshifts, the evolution of elliptical 
and spiral SFHs is followed on the large 1$ \leq z \leq $4 domain,
with different tendencies for the two groups.

\section{Modeling faint galaxy counts}
\subsection{The new version of \textsc{P\'egase}}
The new version of the spectrophotometric model \textsc{P\'egase} (Fioc and
Rocca-Volmerange, 1997, 2000 in preparation) follows the evolution of the stellar energy
distributions (SEDs) for 9 spectral types ranging from ellipticals
to starbursts\footnote{The
codes are accessible on the WEB at
\texttt{http://www.iap.fr/users/fioc} or \texttt{rocca} or by anonymous ftp
at \texttt{ftp.iap.fr} in the directory \texttt{/pub/from$\_$users/pegase}}. 
The main input data are the stellar evolutionary
tracks from the Padova group (Bressan et al., 1994) and the stellar
spectra from Kurucz 1995 corrected by Lejeune et al., 1997.

Exchanges with the interstellar
medium (supernovae ejecta, stellar winds, gas inflow or outflow) are
taken into account and allow to follow the evolution of the
metallicity of the interstellar medium (ISM) and the stellar populations, as
well as the dust opacity.
The extinction is computed by a radiative
transfer code in various geometries corresponding to the morphological
type and for a standard model of grains (Draine \& Lee, 1984).
In disk galaxies, the stars and the dust are distributed homogeneously
in a slab, while in bulges, stars and dust
follow a King-like profile (Fioc and Rocca-Volmerange, 1997). 
The amount of dust is computed from the mass of metals in the ISM.
In the most conservative way, star formation rates are 
proportional to the current gas content, with a type-dependent timescale. 
This timescale, as well as the e-folding time for infall, are estimated
by fitting the synthetic spectra to $z=0$ observational template
spectra and colors.
To this purpose, statistical optical-to-near infrared colors
were recently determined as a function of type, luminosity and
inclination from a catalog of magnitudes corrected for aperture effects
(Fioc and Rocca-Volmerange, 1999b). 
\subsection{Other inputs and results}
The type-dependent $z=0$~luminosity functions of galaxies are
from the Autofib Redshift Survey (Heyl et al., 1997).
The adopted cosmological parameters in the classical formalism 
of Friedmann and Lema\^{\i}tre are
$H_0=65\,\mathrm{km.s^{-1}.Mpc^{-1}}$, while
$\Omega_0$ and $\Lambda_0$
belong to the [0,1] interval. Our two preferred sets
of values of $(\Omega_0; \Lambda_0)$ are $(0.1; 0)$ and $(0.3; 0.7)$.

\begin{figure}
\begin{center}
\epsfig{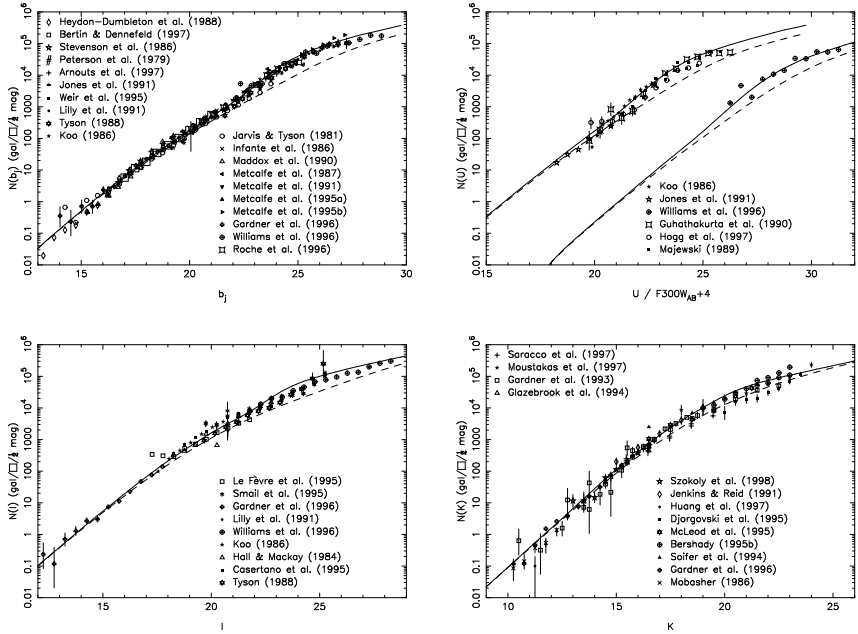}
\end{center}
\caption[]{Predicted galaxy count in
$b_J$, $U$, F300W$_{AB}$, $I$ and $K$ compared to the observations and
normalized to them at $b_J =16$.
The luminosity function is from (Heyl et al., 1997). 
The adopted cosmology corresponds to 
$H_0=65\,\mathrm{km.s^{-1}.Mpc^{-1}}$, 
$\Omega_0=0.3$ and $\Lambda_0=0.7$. The dashed line is
the case without evolution.}
\label{faintcounts}
\end{figure}

\begin{figure}
\begin{center}
\epsfig{figure=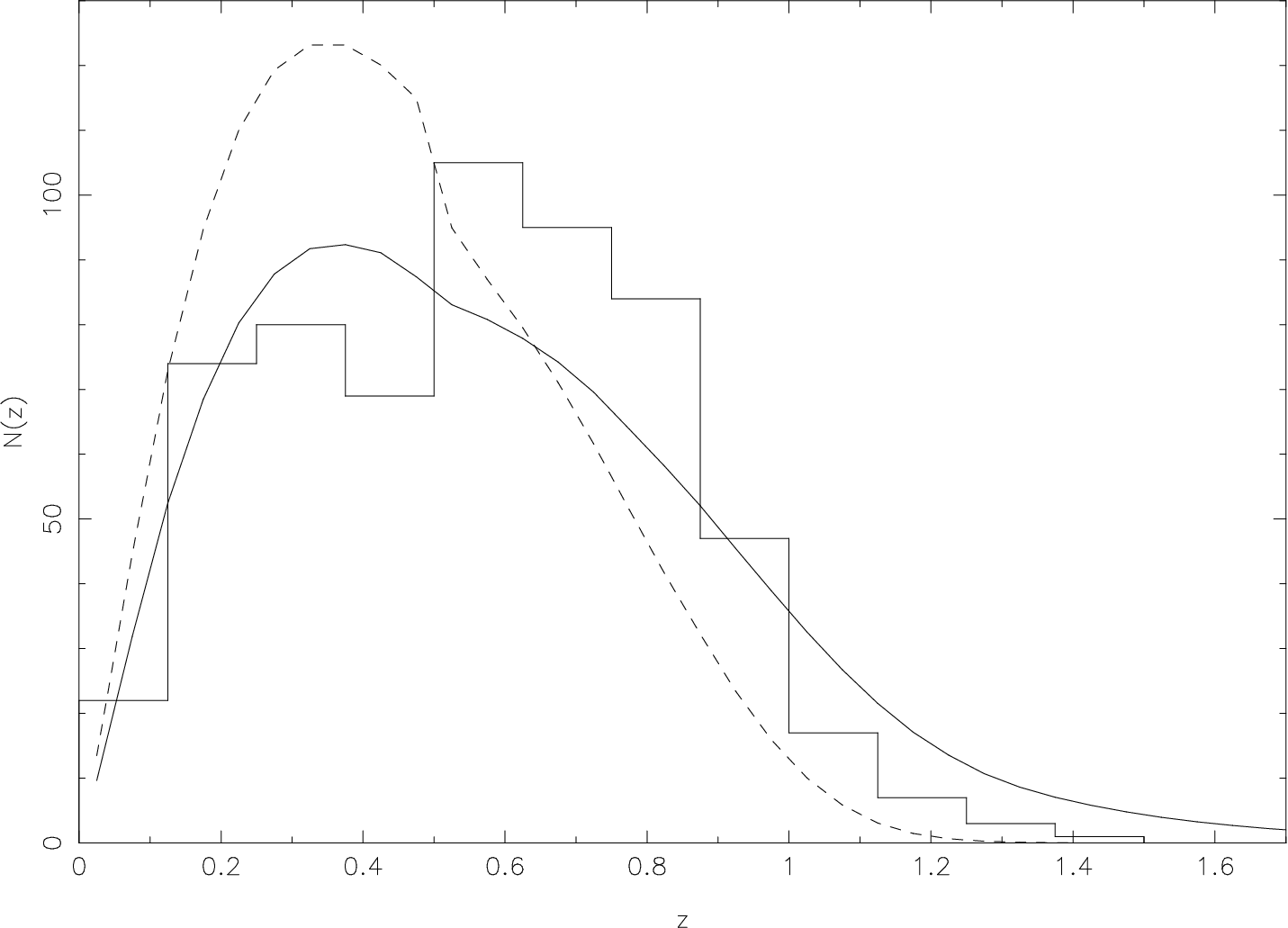,angle=-90,width=0.8\maxfloatwidth}
\label{CFRS-counts}
\epsfig{figure=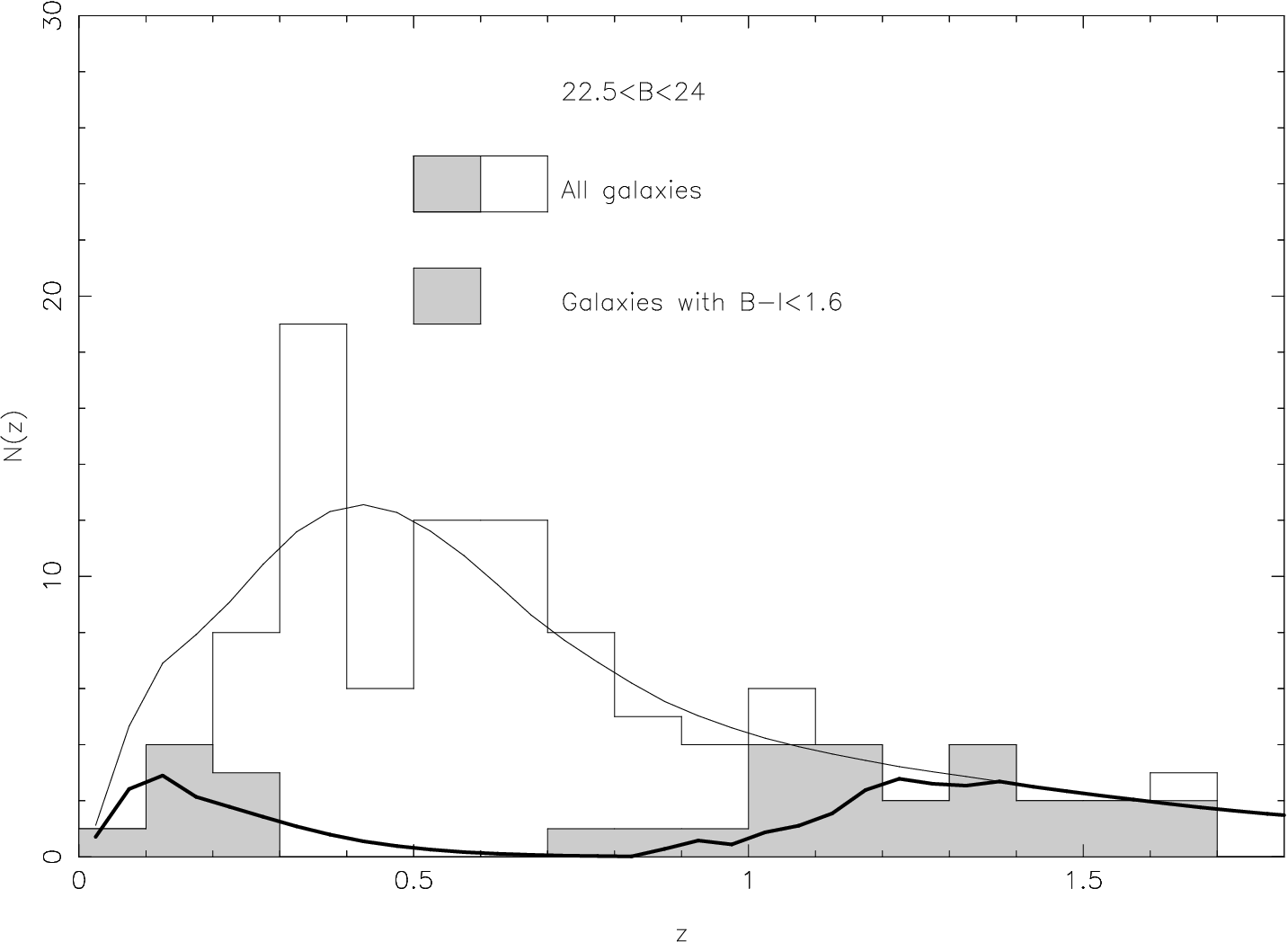,angle=-90,width=0.8\maxfloatwidth}
\label{Cowie-counts}
\end{center}
\caption[]{\textbf{Top:} The redshift distribution N(z) from the CFRS fitted
with our evolution model P\'EGASE. The adopted cosmology corresponds
to H$_0$~=65km.s$^{-1}$.Mpc$^{-1}$, $\Omega_0$=0.3 and
$\Lambda_0$=0.7. Dashed line is the case without evolution.\\ 
\textbf{Bottom:}
The redshift distribution of very faint objects from the Hawaii deep
survey (Cowie et al., 1996) are fitted with our model
P\'EGASE. Cosmological parameters are the same as above. The grey zone identify the bluest (B-I $<$
1.6) populations observed at low and high redshifts and the thick line
is the corresponding prediction.} 
\label{CFRS-Cowie-counts}
\end{figure}

Predicted multiwavelength galaxy counts $N(m)$ and redshift
distributions $N(z)$ are plotted on Figures~\ref{faintcounts}  
and 2
for pure luminosity evolution models and are compared to the
observations. The most contraining data on the cosmology are the
faint counts from the Hubble Deep Field (Williams et al., 1996). 

Because of the relation between $z$ and $t$, redshift distributions
rather put constraints on the SFH, especially
the Hawaii surveys (Cowie et al., 1996, 1999).
The CFRS sample, complete
till $I_{\mathrm{AB}}=22.5$ is well reproduced by our models
(Figure~2, top). However, our SFH 
does not show the rapid evolution found by Lilly et al. (1996)
from the \emph{same} data, which is puzzling.
Figure~2, bottom, highlights
the two populations of blue galaxies (B-I $<$ 1.6) observed
in the Hawaii galaxy survey by Cowie et al. (1996) and in general in 
faint counts. The nearby blue 
population is sufficiently faint ($22.5 \leq B \leq 24)$~
to be assimilated to a population of dwarfs,
undetectable at higher redshifts while the faint distant blue population 
($z > 1$) is also detected in the deepest photometric
surveys as the HDF-N.
\section{Star formation histories}
\begin{figure}
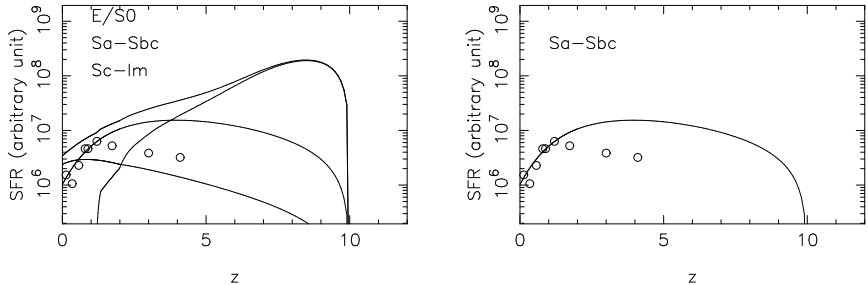

\begin{center}
\epsfig{figure=fig.ps,angle=-90,width=0.45\maxfloatwidth}\qquad
\epsfig{figure=figspir.ps,angle=-90,width=0.45\maxfloatwidth}
\end{center}
\label{Global-Sp-SFH}
\caption{$Left$: The global SFH versus $z$~ is derived from multispectral 
faint counts fits (thick line), the respective contribution
of the three labeled groups are identified (thin lines). Sc-Im
is dominant at $z=0$~and E/SO at high $z$. $z_{for}$=10. $Right$:
the SFH of evolved spiral galaxies is compared to the Lilly et al, 1996 results
from CFRS (empty circles)
}
\end{figure}

For each galaxy type, the star formation law (rate and initial mass
function) determines the spectral evolution. 
Cosmological and evolutionary corrections may then be derived from the
synthetic spectra and be applied to the local type-dependent 
luminosity functions to compute more distant ones and to predict
galaxy counts.
The interest of this procedure is that the models can be compared
directly to the observations, without the inconsistencies brought by
the use of different conversion factors, depending on the wavelength
and the redshift.

Figure~\ref{Global-Sp-SFH} shows the global SFH and the contributions of 
the three groups. The main results are the slow evolution of the
global SFH ($SFH\propto (1+z)^{1.2}$), 
close to Cowie et al., 1999's result, and the six-fold decrease of
the SFR of evolved spirals from $z=1$ to $0$.
No peak is predicted at $z\approx 1$. 
The maximal SFR actually occurs just after the formation of E and S0
galaxies, at a redshift $z_{\mathrm{for}}\approx 10$ on
Figure~\ref{Global-Sp-SFH}.
A similar result (with sharper slopes) would be obtained with
$z_{\mathrm{for}}=4$; such low value is however in worse agreement 
with the data than $z_{\mathrm{for}}=10$~
(Rocca-Volmerange and Fioc, 1999). Yet, the SFR of spiral galaxies
flattens at $z>2$. At such 
high $z$, results are much more sensitive to cosmology and
evolution scenarios than for $z \leq 1$. In particular, the global 
SFR might follow a similar trend if
E/S0 galaxies formed by merging rather than by monolithic collapse.

The examination of the SFH of the various types may help to solve
the contradiction between the ``Madau''
diagram and Cowie et al.'s results. Each galaxy group has a different 
contribution to the global SFH, which must be analyzed separately:
\begin{itemize}
\item[i)]The late-type
group dominates the SFR at $z=0$ (70\%) 
but its weight rapidly decreases at 
$z>1$. Many of these galaxies are starbursts; they
correspond to the very ``blue'' galaxies observed in the UV by
FOCA2000 and provide an explanation for the high local SFR determined 
by Treyer et al. (1998). They also observed in the deeper Cowie et
al. (1999) samples and explain the shallower slope found by the authors. 
\item[ii)]Surprisingly, the SFR of the intermediate-type class
increases by a factor six between $z=0$ and $1$,
not far from the results 
of most bright surveys(CFRS,
H$\alpha$, FOCA2000, ISO/Deep survey). These results explain why the
redshift distribution of the CFRS is well reproduced by our models
(Figure~2, top) and could be due to the fact that
the observed nearby selected samples (CFRS, H$\alpha$ and ISO) are
biased towards bright spirals. 
\item[iii)]The SFR of ellipticals and lenticulars evolves rapidly 
at $z>1$ or $2$, which is consistent with the blue population
of spheroidals discovered in the Hubble Deep Field.
\end{itemize}
By comparison with previous determinations of the global SFH,
our analysis avoid two problems which could  explain 
the strong difference  between Lilly et al., 1996's and
Cowie et al., 1999's results. 
First, the conversion factors used
to convert the observational tracers to star formation rates
(I$_{AB}$ to 2800\AA, H$\alpha$ emission lines, far-IR) 
are very uncertain, mainly because of
the extinction correction. Secund, every sample
suffers from detection limits. Magnitude-limited samples miss
nearby faint galaxies while surface brightness-limited samples do not detect
high-redshift galaxies because of the $(1+z)^4$-fading. 
This argument, already mentioned by Cowie et al., 1999, is partly 
avoided when models are fitted on the deepest surveys.

Another interesting point concerns predictions for spirals at $1 \leq
z \leq 4$. The $SFR$ vs. $z$ curve is flat, in agreement with ISOCAM and
ISOPHOT results (Aussel et al., 1999; Lagache et al., this conference)
after correction of dust effects.

To summarize, the results on a shallow slope of SFH for $z \leq 1$ 
are robust. At higher redshifts, 
a strong increase of star formation rate is predicted
if evolution scenarios of elliptical galaxies are monolithic. However
more uncertainties on cosmological parameters
or high-$z$~star formation processes as merging,
require deeper observations and more detailed analyses 
at high redshift galaxies, depending on metal, dust 
and other parameters.
Implications on the ionisation at high redshifts, the population
of primeval galaxies, all 
observational programs of the future NGST are strong. 

\end{article}
\end{document}